\documentclass[a4,12pt]{article}
\usepackage{amsthm}
\usepackage{amsmath}
\usepackage{amssymb}
\usepackage{natbib}
\usepackage{booktabs} 
\usepackage{fullpage}
\usepackage{xcolor}
\usepackage{enumitem}

\newif\ifpdf
\ifx\pdfoutput\undefined
  \pdffalse                     
  \usepackage[dvips]{graphicx}
\else
  \pdfoutput=1                  
  \pdftrue
  \usepackage[pdftex]{graphicx}
\fi
  
\usepackage[scaled]{helvet}
\usepackage[T1]{fontenc}

\newcommand{\be}{\begin{equation}}
\newcommand{\ee}{\end{equation}}
\newcommand{\bi}{\begin{itemize}}
\newcommand{\ei}{\end{itemize}}
\newcommand{\bea}{\begin{eqnarray}}
\newcommand{\eea}{\end{eqnarray}}

\newcommand{\bfmu}{\boldsymbol{\mu}}

\newcommand{\bfx}{\mathbf{x}}

\newcommand{\bbE}{\mathbb{E}}

\newcommand{\bbR}{\mathbb{R}}

\newcommand{\cut}[1]{}

\newcommand{\tlam}{\tilde{\lambda}}

\title{The Elliptical Quartic Exponential Distribution: An Annular
  Distribution Obtained via Maximum Entropy}

\author{Christopher K I Williams \\ School of Informatics, University
  of Edinburgh, UK}

\date{\today}

\begin{document}
\maketitle

\begin{abstract}
  This paper describes the Elliptical Quartic Exponential distribution
  in $\bbR^D$,
  obtained via a maximum entropy construction by imposing second
  and fourth moment constraints. I discuss relationships to related
  work, analytical expressions for the normalization constant and 
  the entropy, and the conditional and marginal distributions.
\end{abstract}

\setcounter{section}{1}

The maximum entropy construction allows the specification of a
probability distribution in terms of constraints, see e.g.,
\citet[ch.\ 12]{cover-thomas-91}.
Consider a radially symmetric zero-mean distribution in $\bbR^D$ with $r^2 =
\bfx^T \bfx$. Constraints are imposed on the variance $\bbE[r^2] = c_2$ and 
on $\bbE[r^4] = c_4$; the latter implies a constraint
on the ``variance of the variance''. The intuition behind the
construction here is that if this ``variance of the variance'' is small
then  the distribution should be similar to an annulus at some radius
$R$. We define an annular distribution to be one where the
distribution as a function of $r$ is unimodal with the mode away from zero.

The maximum entropy construction gives
\begin{equation}
  p(\bfx) = \frac{1}{Z_D(\lambda_1, \lambda_2)} \exp \left( \lambda_1
  \bfx^T \bfx - \lambda_2  (\bfx^T \bfx)^2   \right), \label{eq:annular}
\end{equation}  
where $Z_D(\lambda_1, \lambda_2)$ is the normalization constant in
$\bbR^D$. We require $\lambda_2 > 0$ so that the distribution is
normalizable. $\lambda_1 > 0$ produces an
annular distribution, while for $\lambda_1\le 0$ the density
decays monotonically from $r=0$.

\begin{figure*}[t]
\centering
\includegraphics[width=4in]{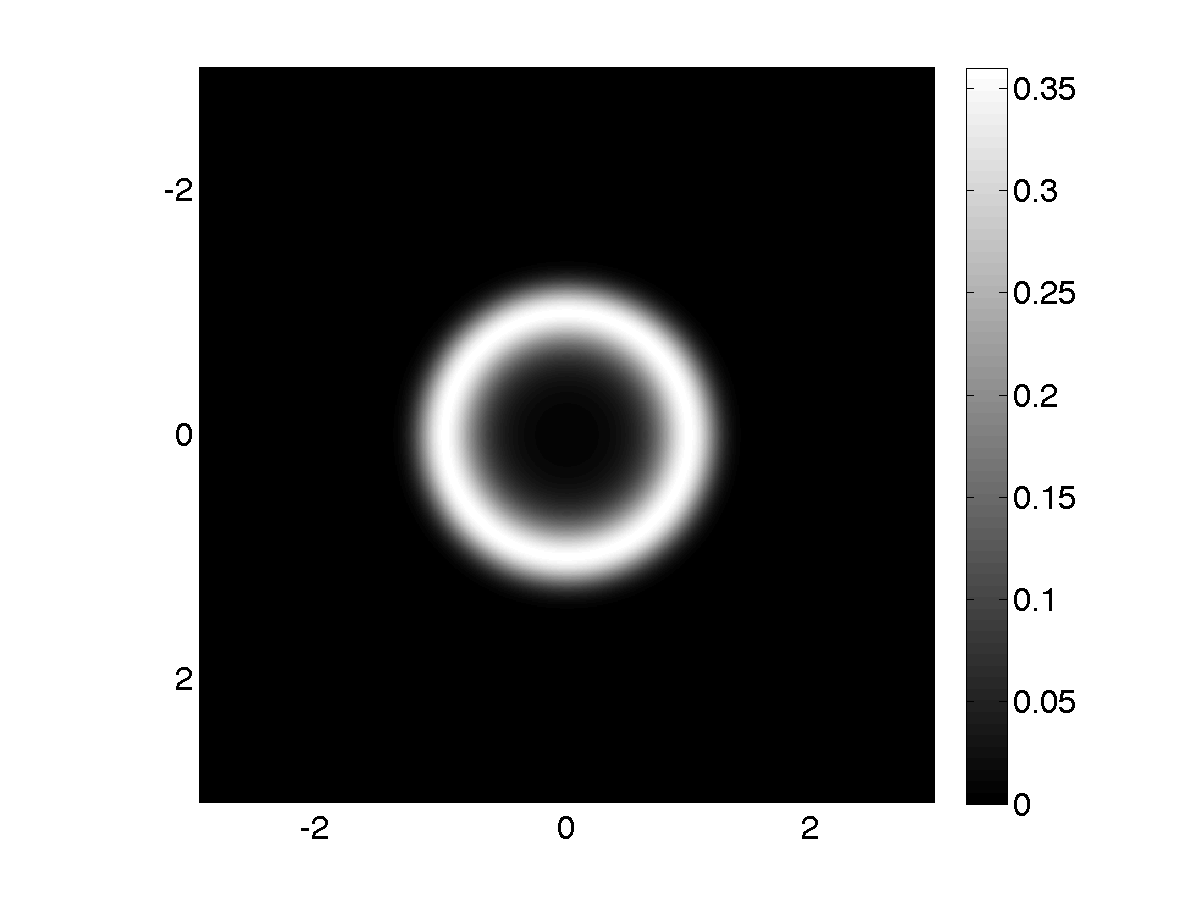}
\caption{A plot of the annular distribution in 2D for
  $R=1$ and $\alpha = 8$.
 \label{fig:ann_plot}}
\end{figure*}

Consider the exponential term in eq.\ \ref{eq:annular} as a
function of $r$ for $\lambda_1 > 0$.
By differentiation we have at the maximum that
$2 \lambda_1 r - 4 \lambda_2 r^3 = 0$. Let the value of $r$ at which
the maximum is reached be denoted by $R$. Hence $\lambda_2 =
\lambda_1/(2 R^2)$. By setting $\lambda_1 = \alpha/R^2$ for $\alpha
>0$, we have
\begin{equation}
  p(\bfx) = \frac{1}{Z_D(\alpha, R)} \exp \alpha \left( 
  \frac{\bfx^T \bfx}{R^2}  - \frac{(\bfx^T \bfx)^2}{2 R^4} \right) .
  \label{eq:annularR}
\end{equation}  
As $\alpha$ increases the thickness of the ring decreases.
A plot of $p(\bfx)$ in 2D with $R=1$ and $\alpha = 8$ is shown in
Figure \ref{fig:ann_plot}.

The distribution can clearly be shifted to a non-zero mean $\bfmu$ and
$\bfx^T \bfx$ can transformed to $\bfx^T \Sigma^{-1} \bfx$ for some
SPD matrix $\Sigma$. I term the distribution in eq.\ \ref{eq:annular}
under this transformation the \emph{Elliptical Quartic Exponential}
distribution, by analogy with the Elliptical Gamma distribution
discussed below; both have elliptical contours.

Below I discuss related work, analytical expressions for
the normalization constant and the entropy, and the
conditional and marginal distributions of the Elliptical Quartic
Exponential distribution.

\subsection{Related work}
\citet{fisher-22} discussed univariate probability densities having
the form $p(x) \propto \exp(-Q_k(x))$, where $Q_k(x) = \sum_{q=1}^k
\alpha_q x^q$, with $k$ even and $\alpha_k > 0$. \citet{matz-78}
discusses the case with $k=4$ known as the \emph{quartic exponential
distribution}, and the special case of $p(x) \propto \exp( - \beta x^2
 - \gamma x^4)$ with $\gamma > 0$ and $\beta$ unrestricted in sign;
 this is termed the symmetric quartic exponential distribution.
The  quartic exponential distribution can be obtained via 
maximum entropy considerations given the first four moments, see e.g.,
\citet{zellner-highfield-88}.

In the multivariate case, \citet{urzua-87} considers $p(\bfx) \propto
\exp(-Q(\bfx))$. If $Q(\bfx)$ is a polynomial of degree $k$ in $D$
dimensions, it can be written as $Q(\bfx) = \sum_{q=1}^k
Q^{(q)}(\bfx)$, where each $Q^{(q)}(\bfx)$ is a homogeneous polynomial of
degree $q$, i.e.\
\begin{equation}
Q^{(q)}(\bfx) = \sum \alpha^{(q)}_{j_1 \ldots j_D} \prod_{i=1}^D
x^{j_i}_i ,
\end{equation}
with the summation taken over all non-negative integer $D$-tuples
$(j_1,\ldots,j_D)$ such that $j_1 + \ldots + j_D = q$.  This is
known as the multivariate quartic exponential distribution. The
maximum possible number of parameters is determined as
$(D+k)!/(D! k!) -1$. However, to
my knowledge, the specific distribution in eq.\ \ref{eq:annular} derived from
the constraints on $r^2$ and $r^4$ only has not been discussed before
in
the literature.

\citet{abramov-10} discusses numerical methods for obtaining the
maximum entropy distribution given moment constraints.

Another route to defining an annular distribution is to first
define an distribution on $r$ which is unimodal with its mode away
from zero, and then to distribute that mass between $r$ and $r+dr$
over the spherical shell at this radius. Using this construction
with the Gamma distribution gives rise to the 
Elliptical Gamma distribution (\citealt*{koutras-86}; see also
\citealt*{sra-hosseini-theis-bethge-15}), defined as
\begin{align}
p_{EG}(\bfx; \Sigma, a,b) & = \frac{\Gamma(D/2)}{\pi^{D/2} \Gamma(a)
b^a |\Sigma|^{1/2}} \phi(\bfx^T \Sigma^{-1} \bfx), 
\end{align}
where $\phi(t) = t^{a - D/2} e^{-t/b}$,
$a, \, b > 0$ and $\Sigma$ is a symmetric positive definite (SPD) matrix.
Differentiation of $\phi(t)$ for $t = r^2 = \bfx^T \Sigma^{-1} \bfx$ 
shows that it reaches a maximum at $r^2 = b(a - D/2)$ for 
$a > D/2$, giving rise to an annular distribution quite similar
to the Elliptical Quartic Exponential distribution.

\subsection{Normalization constant}
A remaining issue is to obtain an analytic expression for 
$Z_D(\lambda_1, \lambda_2)$. We have that 
\begin{equation}
  Z_D(\lambda_1, \lambda_2) = \int_0^{\infty} S_{D-1} r^{D-1}  \; e^{(\lambda_1
  r^2 - \lambda_2 r^4)} dr. 
\end{equation}
where $S_n$ denotes the surface area of the unit sphere in
$n$-dimensions; for example the unit 1-sphere is the unit circle in
$\bbR^2$, so $S_1 = 2 \pi$.
Now consider the change of variable $y = r^2$, with $dy = 2 r dr$. Hence
\begin{equation}
  Z_D(\lambda_1, \lambda_2) = \frac{S_{D-1}}{2}
  \int_0^{\infty} y^{D/2 -1} e^{(\lambda_1 y -\lambda_2 y^2)} dy. 
\end{equation}

\citet{gradshteyn-ryzhik-07} equation 3.462.1 is
\begin{equation}
\int_0^{\infty} y^{\nu-1} e^{- \beta y^2 - \gamma y} dy =
(2 \beta)^{-\nu/2} \Gamma(\nu) \exp(\frac{\gamma^2}{8 \beta})
D_{-\nu}(\frac{\gamma}{\sqrt{2 \beta}}) , \label{eq:gr3.462.1}
\end{equation}
where $D_{\nu}$ is a Parabolic Cylinder Function \citep[9.24-9.25]{gradshteyn-ryzhik-07}.
Hence by identifying $\nu = D/2$, $\beta = \lambda_2$ and $\gamma = -
\lambda_1$ we have that 
\begin{equation}
  Z_D(\lambda_1, \lambda_2) = \frac{S_{D-1}}{2} 
(2 \lambda_2)^{-D/4} \Gamma(D/2) \exp(\frac{\lambda_1^2}{8 \lambda_2})
D_{-D/2}(\frac{- \lambda_1}{\sqrt{2 \lambda_2}}) . \label{eq:Z_D}
\end{equation}

For $D=2$ or $\nu=1$ using the relationship $D_{-1}(z) =
\sqrt{\frac{\pi}{2}} e^{z^2/4}  [1 - \Phi(\frac{z}{\sqrt{2}})]$ from
\citet[9.254.1]{gradshteyn-ryzhik-07}, where
$\Phi(\cdot)$ denotes the error function (erf)
$\Phi(x) = \frac{2}{\sqrt{\pi}} \int_0^x e^{-t^2} dt$, we have
\begin{equation}
Z_2(\lambda_1, \lambda_2) = \frac{\pi}{2} \sqrt{\frac{\pi}{\lambda_2}}
\exp \left( \frac{\lambda_1^2}{4 \lambda_2} \right) \left[ 1 - \Phi(\frac{-\lambda_1}{2
    \sqrt{\lambda_2}}) \right] .
\end{equation}

In eq.\ \ref{eq:Z_D} a general expression for $Z_D(\lambda_1, \lambda_2)$ is given
in terms of the Parabolic Cylinder Function $D_{\nu}$. It is of
interest to see if this can be expressed in terms of more familiar
functions for the case of $D=1$. As discussed in \citet[sec.\ 2]{matz-78},
the cases of $\lambda_1 \gtrless 0$ are considered separately.
\citet[pp.\ 5-9]{otoole-33} discusses the case for $\lambda_1 > 0$, and
obtains a complicated expression involving the Bessel functions
$J_{1/4}$ and $J_{-1/4}$. 
A series expansion  (eq.\ 2) is also given which is recommended for
computation. For $\lambda_1 < 0$, the integral 3.323.3 in
\citet{gradshteyn-ryzhik-07}  results in an expression
involving the modified Bessel function $K_{1/4}$.
Of course numerical quadrature is straightforward for the 1D case.

\subsection{Entropy}
We have that $- \log p(\bfx) = \lambda_2 (\bfx^T \bfx)^2 -  \lambda_1
(\bfx^T \bfx) + \log Z_D(\lambda_1, \lambda_2)$, and hence that the
entropy is given by
\begin{equation}
  H(p) = \bbE_p[- \log p(\bfx)] = \lambda_2 \bbE_p[r^4] - \lambda_1 \bbE_p[r^2] + \log Z_D(\lambda_1, \lambda_2).
\end{equation}
In general we have that
\begin{equation}
\bbE_p[r^k]   = S_{D-1} \int_0^{\infty} r^{k+D-1} e^{(\lambda_1
  r^2 - \lambda_2 r^4)} dr. 
\end{equation}  
By using the change of variable $y = r^2$, this integral can be
evaluated as per eq.\ \ref{eq:gr3.462.1} for $k=2$ and $k=4$, and
hence the entropy can be computed.

\subsection{Conditional distribution}
Let $p(\bfx) \propto
e^{J(\bfx)}$, where $J(\bfx) = \lambda_1  \bfx^T \bfx - \lambda_2  (\bfx^T \bfx)^2$.
Now consider splitting $\bfx$ into two parts $\bfx =
(\bfx_1^T,\bfx_2^T)^T$, so that $\bfx^T \bfx = \bfx_1^T \bfx_1 + \bfx_2^T \bfx_2$.
Hence 
\begin{equation}
  J(\bfx_1, \bfx_2) = \lambda_1  (\bfx_1^T \bfx_1 + \bfx_2^T \bfx_2)
  - \lambda_2  \left( (\bfx_1^T \bfx_1)^2 + (\bfx_2^T \bfx_2)^4 + 2 (\bfx_1^T \bfx_1)(\bfx_2^T \bfx_2)   \right) .
\end{equation}
Consider the conditional distribution $p(\bfx_1|\bfx_2) \propto
p(\bfx_1,\bfx_2)$ when keeping $\bfx_2$ fixed. Hence we obtain
\begin{equation}
  p(\bfx_1|\bfx_2) \propto \exp \left[ (\lambda_1 - 2 \lambda_2   (\bfx_2^T \bfx_2)) \bfx_1^T \bfx_1 - \lambda_2 (\bfx_1^T \bfx_1)^2  \right].
\end{equation}  
Let $\lambda'_1 = \lambda_1 - 2 \lambda_2 (\bfx_2^T \bfx_2)$. Recall
that $\lambda_1 = 2R^2 \lambda_2$, hence $\lambda'_1 = \lambda_1 (1 -
(\bfx_2^T \bfx_2)/R^2)$. If $(\bfx_2^T \bfx_2) < R^2$ then $\lambda'_1
> 0$ and the conditional distribution is an annular distribution. But
if $(\bfx_2^T \bfx_2) \ge R^2$ then $\lambda'_1 \le 0$ the conditional
distribution is a unimodal distribution centered at the origin.

For a geometric intuition, consider an annular distribution in $D=3$
dimensions, and condition on $x_3 = c$ to obtain a
2D slice through the 3D dimension. If $c < R$ then this is an annular
distribution, while for $c\ge R$ it is a
unimodal distribution centered at the origin.

\subsection{Marginal distribution}
For simplicity we consider the 2D distribution with
$\bfx^T \bfx = x_1^2 + x_2^2$. Then 
\begin{equation}
p(x_1,x_2) =  \frac{1}{Z_2(\lambda_1, \lambda_2)} \exp \left( \lambda_1
  (x_1^2 + x_2^2)  - \lambda_2  (x_1^4 + 2 x_1^2 x_2^2 + x_2^4)   \right). \label{eq:annular2D}
\end{equation}  
Thus
\begin{align}
  p(x_1) &=    \int_{-\infty}^{\infty}  p(x_1, x_2) dx_2 \\
         &=  \frac{1}{Z_2(\lambda_1, \lambda_2)} \exp (\lambda_1 x_1^2
    - \lambda_2 x_1^4)  \int_{-\infty}^{\infty}
\exp ((\lambda_1 - 2 \lambda_2 x_1^2) x_2^2  - \lambda_2 x_2^4) dx_2
. \label{eq:margx1}
\end{align}
Let $\tlam_1 = (\lambda_1 - 2 \lambda_2 x_1^2)$. Observe that 
the integral above is equal $Z_1(\tlam_1,\lambda_2)$.
Recalling that $\tlam_1 = (\lambda_1 - 2 \lambda_2 x_1^2)$ is a
function of $x_1$, the expression for $Z_1(\tlam_1,\lambda_2)$
can be combined with the factors
before the integral in eq.\  \ref{eq:margx1} to obtain the marginal
distribution.
A plot of the marginal distribution for the case shown in Figure \ref{fig:ann_plot}
shows that it is bimodal, with peaks just inside $\pm R$. 

This marginalization can be extended to multivariate $\bfx_1, \,
\bfx_2$ by replacing $x^2_1$ by $\bfx_1^T \bfx_1$, and similarly for
$\bfx_2$. However, note that in the integration in
eq.\ \ref{eq:margx1}
there will be a factor of $S_{D_2 -1} r_2^{D_2 -1}$ coming in when
changing coordinates to $r_2^2 = \bfx^T_2 \bfx_2$,
corresponding to $Z_{D_2}(\tlam_1,\lambda_2)$.

\subsection*{Acknowledgments}
I thank my colleagues Iain Murray, Michael Gutmann and Natalia
Bochkina and an anonymous referee for helpful comments.

\bibliographystyle{apalike}
\bibliography{refs}

\end{document}